\documentclass[aps,superscriptaddress,twocolumn,showpacs,preprintnumbers,amsmath,amssymb,showkeys,textcomp]{revtex4}
\usepackage[cp1251]{inputenc}
\usepackage[english]{babel}
\usepackage{textcomp,amssymb,amsmath}
\usepackage{amstext,textcomp} 
\usepackage[dvips]{hyperref}
\usepackage[mathscr]{eucal}
\usepackage{longtable}
\usepackage{graphicx}
\usepackage[english]{babel}

\begin{document}
\title{Analytical Solution of Mathieu Equation}

\author{Dmitri Yerchuck (a), Alla Dovlatova (b), Yauhen Yerchak(c), Felix Borovik (a)  \\
(a) - Heat-Mass Transfer Institute of National Academy of Sciences of RB, Brovka Str., 15,  Minsk, 220072,  RB \\(b) - M.V.Lomonosov Moscow State University, Moscow, 119899\\
(c) Belarusian State University, Nezavisimosti Avenue 4, Minsk, 220030, RB}
\date{\today}
             
\begin{abstract}  
The general solution of the homogeneous damped Mathieu equation in the analytical form, allowing its practical using in many applications, including superconductivity studies, without numerical calculations has been found.
\end{abstract}

                             
\maketitle
\section{Introduction and Background}

It is known, that a number of physical phenomena can be described mathematically by  
Mathieu equation. For instance,
the equation
of motion for the flux lattice in the  supeconductor theory  is the following \cite{Kim}

\begin{equation}
\begin{split}
\label{Eq1}
 m\frac{d^2y(t)}{dt^2} + \eta\frac{dy(t)}{dt} + (K_0 + k \cos\omega{t})y(t) = \frac{BJ_0}{c}\cos\Omega{t},
\end{split}
\end{equation}

where $B$ is the magnetic induction in the sample, $ m$ is the total mass (per unit length) of the
flux lattice, $\eta$ is the viscosity coefficient, $\omega$ is the frequency of the modulating magnetic field, $c$ is the light velocity, $J_0$ and
$\Omega$ are the amplitude and the frequency of the microwave
current, respectively, $K_0$ and  $k$ are amplitudes of  constant
component  and an  alternating component in the function $K(t)$ in the relationship

\begin{equation}
\begin{split}
\label{Eq2}
F(t) = -K(t)y(t)
\end{split}
\end{equation}

between the  force $F(t)$ and small displacement $y(t)$ of the flux lattice from its equilibrium
position, that is

\begin{equation}
\label{Eq3}
K(t) = K_0 + k \cos\omega{t}.
\end{equation}

The equation (\ref{Eq1})
is the inhomogeneous damped Mathieu equation.

Mathieu equation is well knowm in the theory of differential equations, see for example \cite{Kotowski}, \cite{Meixner}, \cite{Bateman}, \cite{Lachlan}.
At that,the so-called general Mathieu equation is the following equation

\begin{equation}
\begin{split}
\label{Eq4}
 \frac{d^2y(t)}{dt^2} + [h - 2\theta \cos({2t})] y(t) = 0,
\end{split}
\end{equation}

where $h$ and $\theta$ are real or complex constants. The known solution of the general Mathieu equation (\ref{Eq4}) is built in the form

\begin{equation}
\begin{split}
\label{Eq5}
y(t) = \exp({\mu t}) P(t),
\end{split}
\end{equation}

where $P(t)$ is a periodical function with the period, equal to $\pi$, $\mu$ is so-called characteristic  index, depending on the values of $h$ and $\theta$. The function
 
\begin{equation}
\begin{split}
\label{Eq6}
y(t) = \exp({-\mu t}) P(-t),
\end{split}
\end{equation}

represents itself the second solution. The solutions (\ref{Eq5}) and (\ref{Eq6}) are linearly independent, they produce the fundamental system of the solutions, excluding the case, when $i\mu \in Z$, that is, to the set of whole numbers \cite{Bateman}. Further, the solution (\ref{Eq5}) is written formally in the form of the following infinite series

\begin{equation}
\begin{split}
\label{Eq7}
y(t) = \sum_{n = -\infty}^{\infty}c_n\exp{(\mu + 2ni) t}, 
\end{split}
\end{equation}

in which for the set of  coefficients  $\{c_n\}$ the following recurrent relations

\begin{equation}
\begin{split}
\label{Eq8}
-\theta c_{n-1} +[h + (\mu + 2ni)^2] c_{n}-\theta c_{n+1} = 0, n \in Z
\end{split}
\end{equation}

were obtained.
It is the only algorithm for the numerical solution. In fact, the analytical solution of the general Mathieu equation (\ref{Eq4}) was not found. The existence the only algorithn for the numerical solution is inconveniently for the  practical usage of Mathieu equation, especially in the cases of physical applications when 
analytical dependences are required to understand the physical processes. For instance, the authors of the work \cite{Kim}  have preferred instead of  trying to solve  the equation (\ref{Eq1}) numerically to use  the linearized equation

\begin{equation}
\begin{split}
\label{Eq9}
 m\frac{d^2\delta y}{dt^2} + \eta{d\delta y}{dt} + K_0\delta y = - k \cos\omega{t}y_0,
\end{split}
\end{equation}

 in which $y_0$ is the particular solution of
(\ref{Eq1}) for the case of $k = 0$. Let us remark, that an (in)homogeneous damped  Mathieu equation (\ref{Eq4}) can be reduced to an (in)homogeneous general Mathieu equation by the following transformation of  the  function $y(t)$: $y(t) = w(t)\exp(-\frac {1}{2}\eta t)$.

The authors of the work \cite{Kim} have found, to the first order in $k$, $\eta$, and $\omega$, that the value of an electric field $E(t)$ induced is

\begin{equation}
\begin{split}
\label{Eq10}
E(t) = \frac{B^2J_0\Omega}{|K_0|c^2} [1 - \epsilon \cos(\omega{t} - \phi)]  \sin(\Omega{t} -\alpha),
\end{split}
\end{equation}

where $\epsilon = \frac{k}{K_0}$, $\tan\phi =  \frac{2\eta\omega}{K_0}$, and $\tan\alpha =  \frac{\eta\Omega}{K_0}$. So, the equation  (\ref{Eq10}) indicates that the induced $E(t)$ field is
 amplitude modulated.
In other words, the concrete physical mechanism and its analytical description were established, alhough for rather restricted ranges of the parameters used. Really, the
equation (\ref{Eq10}) is valid for $k \ll |K_0|$, $\omega \ll \Omega$, and
$|K_0|\gg m\Omega^2$. 
 At the same time, the authors of  \cite{Kim} remark, that
when static magnetic field $H$ becomes too large [larger than 800 G], the flux
structure becomes probably too complex for the simple model proposed to remain valid. 

It is clear on the given example, that an analytical solution of Mathieu equation remains to be very actual for its applications in physical sciences and in engineering.

On the other hand, an analytical solution of Mathieu equation has also the mathematical theoretical aspect. It is determined by the fact that the solution of a number of differential equations is reduced to the solution of Mathieu equation. They, for example, are

1)

\begin{equation}
\begin{split}
\label{Eq11}
(1-t^2)\frac{d^2y(t)}{dt^2} + t\frac{dy(t)}{dt} + (2a t^2  + b)y(t)  = 0,
\end{split}
\end{equation}

The transformation of variable $t = cos z$ leads to the Mathieu equation

\begin{equation}
\begin{split}
\label{Eq12}
\frac{d^2y(z)}{dz^2}  + (a  + b + a cos2z)y(z)  = 0,
\end{split}
\end{equation}

2)

\begin{equation}
\begin{split}
\label{Eq13}
2t(t-1)\frac{d^2y(t)}{dt^2} + (2t -1)\frac{dy(t)}{dt} + (a t  + b) y(t)  = 0,
\end{split}
\end{equation}

The transformation of variable $t = cos^2 z$ leads to the Mathieu equation

\begin{equation}
\begin{split}
\label{Eq14}
\frac{d^2y(z)}{dz^2}  - (a  + 2b + a cos2z)y(z)  = 0,
\end{split}
\end{equation}

3)

\begin{equation}
\begin{split}
\label{Eq15}
\frac{d^2y(t)}{dt^2}  + (a\sin \lambda t  + b) y(t)  = 0,
\end{split}
\end{equation}

The transformation of variable $\lambda t = 2z + \frac{\pi}{2}$ leads to the Mathieu equation

\begin{equation}
\begin{split}
\label{Eq16}
\frac{d^2y(z)}{dz^2}  - (\frac{4b}{\lambda^2} +  \frac{4a}{\lambda^2}cos2z) y(z)  = 0,
\end{split}
\end{equation}

4)

The equations 

\begin{equation}
\begin{split}
\label{Eq17}
&\frac{d^2y(t)}{dt^2} +  (a\sin^2t  + b) y(t)  = 0,\\
&\frac{d^2y(t)}{dt^2} +  (a\cos^2t  + b) y(t)  = 0
\end{split}
\end{equation}

are transformed to  Mathieu equations by using of trigonometric formulae

\begin{equation}
\begin{split}
\label{Eq18}
&2\sin ^2 t = 1 - \cos 2t,\\
&2\cos ^2 t = 1 + \cos 2t
\end{split}
\end{equation}

correspondingly.

The aim of the given work is to find the analytical solution of the general Mathieu equation.

\section{Results}

\textbf{Theorem.} The general solution of the general Mathieu equation can be represented analytically to be the superposition of Bessel functions of the first and the second kinds.

\textbf{Proof.}

It is evident, that the equation (\ref{Eq1}) can be represented in the form

\begin{equation}
\begin{split}
\label{Eq19}
&\frac{1}{2} m\frac{d^2y}{dt^2} + \frac{1}{2}\eta\frac{dy}{dt} + \frac{1}{2}K_0 y + \frac{1}{2}k \exp( {i\omega{t}}) y - \frac{BJ_0}{2c}\cos\Omega{t} +\\
&[\frac{1}{2} m\frac{d^2y}{dt^2} + \frac{1}{2}\eta\frac{dy}{dt} + \frac{1}{2}K_0 y + \frac{1}{2}k \exp(-{i\omega{t}}) y -\\
& \frac{BJ_0}{2c}\cos\Omega{t}] = (\hat{L}_1 +\hat{L}_2) y = 0,
\end{split}
\end{equation}

where $\hat{L}_1$ and $\hat{L}_2$ are differential operators

\begin{equation}
\begin{split}
\label{Eq20}
&\hat{L}_1 = \frac{1}{2} m\frac{d^2}{dt^2} + \frac{1}{2}\eta\frac{d}{dt} + \frac{1}{2}K_0  + \frac{1}{2}k \exp( {i\omega{t}})  - \frac{BJ_0}{2c}\cos\Omega{t} \\
&\hat{L}_2 = [\frac{1}{2} m\frac{d^2}{dt^2} + \frac{1}{2}\eta\frac{d}{dt} + \frac{1}{2}K_0  + \frac{1}{2}k \exp(-{i\omega{t}})  -\\
& \frac{BJ_0}{2c}\cos\Omega{t}] .
\end{split}
\end{equation}

It is also evident that the partial solution of the starting equation will correspond to the intersection of sets of the solutions satysfying simultaneously to the equations

\begin{equation}
\begin{split}
\label{Eq21}
&\hat{L}_1 y = 0,\\
&\hat{L}_2 y = 0.
\end{split}
\end{equation}

So, we have to solve the equations

\begin{equation}
\begin{split}
\label{Eq22}
&\frac{1}{2} m\frac{d^2y}{dt^2} + \frac{1}{2}\eta\frac{dy}{dt} + \frac{1}{2}K_0 y + \frac{1}{2}k \exp( {i\omega{t}}) y = \\
&\frac{BJ_0}{2c}\cos\Omega{t},\\
&\frac{1}{2} m\frac{d^2y}{dt^2} + \frac{1}{2}\eta\frac{dy}{dt} + \frac{1}{2}K_0 y + \frac{1}{2}k \exp(-{i\omega{t}}) y = \\
&\frac{BJ_0}{2c}\cos\Omega{t} 
\end{split}
\end{equation}

or in the equivalent form

\begin{equation}
\begin{split}
\label{Eq23}
&\frac{d^2y}{dt^2} + \frac{\eta}{m}\frac{dy}{dt} + \frac{K_0}{m} y + \frac{k}{m} \exp({i\omega{t}}) y  =\\
&\frac{BJ_0}{mc}\cos\Omega{t},\\
&\frac{d^2y}{dt^2} + \frac{\eta}{m}\frac{dy}{dt} + \frac{K_0}{m} y + \frac{k}{m} \exp(-{i\omega{t}}) y  =\\
&\frac{BJ_0}{mc}\cos\Omega{t},\\
\end{split}
\end{equation}

The case

\begin{equation}
\begin{split}
\label{Eq24}
&\frac{1}{2} m\frac{d^2y}{dt^2} + \frac{1}{2}\eta\frac{dy}{dt} + \frac{1}{2}K_0 y + \frac{1}{2}k \exp({i\omega{t}}) y - \\
&\frac{BJ_0}{2c}\cos\Omega{t} =\\
&-[\frac{1}{2} m\frac{d^2y}{dt^2} + \frac{1}{2}\eta\frac{dy}{dt} + \frac{1}{2}K_0 y + \frac{1}{2}k \exp(-{i\omega{t}}) y - \\
&\frac{BJ_0}{2c}\cos\Omega{t}]
\end{split}
\end{equation}

can be  also studied. There are a number of variants to simplify the solution of (\ref{Eq24}) using the symmetry of $y(t)$. 
In the case, if $y(t)$ is even, $\frac{dy}{dt}$ is uneven, $\frac{d^2y}{dt^2}$ is even from (\ref{Eq24}) we obtain by $t'\rightarrow -t$

\begin{equation}
\begin{split}
\label{Eq25}
& m\frac{d^2y}{dt'^2} +  K_0 y + \frac{1}{2}k \exp({i\omega{t'}}) y + \frac{1}{2}k \exp({-i\omega{t'}}) y =\\ 
&\frac{BJ_0}{c}\cos\Omega{t'}.
\end{split}
\end{equation}

Given case is equivalent to the task above formulated by the relations  (\ref{Eq19}) - (\ref{Eq21}). Further,
if $y(t)$ is even, $\frac{dy}{dt}$ is uneven, $\frac{d^2y}{dt^2}$  is even from (\ref{Eq24}) we obtain  by $t'\rightarrow -t$

\begin{equation}
\begin{split}
\label{Eq26}
\eta\frac{dy}{dt}  - \frac{BJ_0}{2c}\cos\Omega{t} = 0,\\
\end{split}
\end{equation}

that is the simple differential equation of the first order , the solution of which is evident.

The homogeneous  equations, corresponding to inhomogeneous  equations (\ref{Eq1})
that is the equations

\begin{equation}
\begin{split}
\label{Eq27}
&\frac{d^2y}{dt^2} + \frac{\eta}{m}\frac{dy}{dt} + \frac{K_0}{m} y + \frac{k}{m} \exp({i\omega{t}}) y = 0\\
&\frac{d^2y}{dt^2} + \frac{\eta}{m}\frac{dy}{dt} + \frac{K_0}{m} y + \frac{k}{m} \exp(-{i\omega{t}}) y = 0
\end{split}
\end{equation}
can be solved strictly. Let us designate $\frac{\eta}{m} = a$     $\frac{K_0}{m} = b$,
$\frac{k}{m} = c$,  $i\omega = \lambda$, $-i\omega = \lambda '$.

Then, according to \cite{Zaitsev}  the solutions are
 
\begin{equation}
\begin{split}
\label{Eq28}
&y(t)^{[1]}  = \exp(-\frac{\eta}{2m}t)[C^{[1]}_1 J_\nu(\frac{2\sqrt{\frac{K_0}{m}}}{i\omega}\exp(\frac{i\omega t}{2})) + \\
&C^{[1]}_2 Y_\nu(\frac{2\sqrt{\frac{K_0}{m}}}{i\omega}\exp(\frac{i\omega t}{2}))]
\end{split}
\end{equation}
for the first equation in (\ref{Eq27}) and

\begin{equation}
\begin{split}
\label{Eq29}
&y(t)^{[2]}  = \exp(-\frac{\eta}{2m}t)[C^{[2]}_1 J_{\nu{'}} (\frac{2\sqrt{\frac{K_0}{m}}}{-i\omega}\exp(\frac{-i\omega t}{2})) + \\
&C^{[2]}_2 Y_{\nu{'}}(\frac{2\sqrt{\frac{K_0}{m}}}{-i\omega}\exp(\frac{-i\omega t}{2}))],
\end{split}
\end{equation}

for the second equation in (\ref{Eq27}). Here

\begin{equation}
\begin{split}
\label{Eq30}
&J_\nu(\frac{2\sqrt{\frac{K_0}{m}}}{i\omega}\exp(\frac{i\omega t}{2})),
J_{\nu{'}} (\frac{2\sqrt{\frac{K_0}{m}}}{-i\omega}\exp(\frac{-i\omega t}{2})),\\
&Y_\nu(\frac{2\sqrt{\frac{K_0}{m}}}{i\omega}\exp(\frac{i\omega t}{2})),
Y_{\nu{'}}(\frac{2\sqrt{\frac{K_0}{m}}}{-i\omega}\exp(\frac{-i\omega t}{2}))
\end{split}
\end{equation}
 
are Bessel functions of the first kind and of the second kind of index

\begin{equation}
\begin{split}
\label{Eq31}
\nu = \frac{\sqrt{({\frac{\eta}{m}})^2 -4 \frac{k}{m}}}{i\omega}
\end{split}
\end{equation}

for the equation (\ref{Eq30}) and  of index

\begin{equation}
\begin{split}
\label{Eq32}
\nu{'} = \frac{\sqrt{({\frac{\eta}{m}})^2 -4 \frac{k}{m}}}{-i\omega}
\end{split}
\end{equation}

for the equation (\ref{Eq31}). 

Then the set of functions, satisfying to the relation

\begin{equation}
\begin{split}
\label{Eq33}
y(t)^{[1]}\cap y(t)^{[2]}  
\end{split}
\end{equation}
 
will be the solution of the homogeneous  equation, corresponding to the starting inhomogeneous equation (\ref{Eq1}). Given conclusion is correct, if $y(t)^{[1]}\cap y(t)^{[2]} \neq \varnothing$. In particular, it is satisfied for the case $C^{[1]}_1 = 0$, $C^{[2]}_1 = 0$ and by

 \begin{equation}
\begin{split}
\label{Eq34}
&C^{[1]}_2 Y_\nu(\frac{2\sqrt{\frac{K_0}{m}}}{i\omega}\exp(\frac{i\omega t}{2})) =\\
 &C^{[2]}_2 Y_{-\nu{}}(\frac{2\sqrt{\frac{K_0}{m}}}{-i\omega}\exp(\frac{-i\omega t}{2})),
\end{split}
\end{equation}
 that is, in the case
 
\begin{equation}
\begin{split}
\label{Eq35}
&C^{[1]}_2 \frac {J_\nu(\frac{2\sqrt{\frac{K_0}{m}}}{i\omega}\exp(\frac{i\omega t}{2})) \cos\pi\nu - J_{-\nu}(\frac{2\sqrt{\frac{K_0}{m}}}{-i\omega}\exp(\frac{-i\omega t}{2}))}{\sin\pi\nu} = \\
&C^{[2]}_2  \frac {-J_{-\nu}(\frac{2\sqrt{\frac{K_0}{m}}}{-i\omega}\exp(\frac{-i\omega t}{2})) \cos\pi\nu + J_{\nu}(\frac{2\sqrt{\frac{K_0}{m}}}{i\omega}\exp(\frac{i\omega t}{2}))}{\sin\pi\nu},
\end{split}
\end{equation}

Hence we obtain the conditions, by which  the relation

\begin{equation}
\begin{split}
\label{Eq36}
C^{[1]}_2\exp(-\frac{\eta}{2m}t) Y_\nu(\frac{2\sqrt{\frac{K_0}{m}}}{i\omega}\exp(\frac{i\omega t}{2})) 
\end{split}
\end{equation}

will be the solution of the homogeneous damped Mathieu equation. They are

\begin{equation}
\begin{split}
\label{Eq37}
&\frac{C^{[1]}_2 \cos\pi\nu - C^{[2]}_2}{\sin\pi\nu } = 0,\\
&\frac{C^{[1]}_2\cos\pi\nu - C^{[1]}_2}{\sin\pi\nu} = 0
\end{split}
\end{equation}
that is,  by

\begin{equation}
\begin{split}
\label{Eq38}
&\frac{\cos^2\pi\nu -1}{\sin\pi\nu} = 0,\\
&C^{[1]}_2 \neq 0.
\end{split}
\end{equation}
Consequently,
$\nu \in Z$, that is $\nu$ has to belong to the set of whole numbers.

Therefore, the relation (\ref{Eq36}) is the solution of homogeneous damped Mathieu equation by $\forall C^{[1]}_2 \neq 0 \in C$ and by $\nu \in Z$. It means, that there are restrictions on possible values of parameters in the starting homogeneous damped Mathieu equation. They are the following

\begin{equation}
\begin{split}
\label{Eq39}
\nu = (\frac{\sqrt{({\frac{\eta}{m}})^2 -4 \frac{k}{m}}}{i\omega}) \in Z,
\end{split}
\end{equation}
that is 
\begin{equation}
\begin{split}
\label{Eq40}
\nu = (\frac{\sqrt{{4 \frac{k}{m}-(\frac{\eta}{m}})^2}}{\omega}) \in Z
\end{split}
\end{equation}

It is evident that by $\eta = 0$ we obtain the solution of the homogeneous undamped Mathieu equation, that is the solution of the general Mathieu equation.

 To obtain the fundamental system of the solutions of  the homogeneous damped Mathieu equation, we have to find the second linearly independent solution. It can be easily  done, if to take into account, that
 
\begin{equation}
\begin{split}
\label{Eq41}
&J_{\nu{'}} (\frac{2\sqrt{\frac{K_0}{m}}}{-i\omega}\exp(\frac{-i\omega t}{2})) =
J_{-\nu}(\frac{2\sqrt{\frac{K_0}{m}}}{i\omega}\exp(\frac{i\omega t}{2})) = \\ &(-1)^{\nu} 
J_{\nu{}} (\frac{2\sqrt{\frac{K_0}{m}}}{i\omega}\exp(\frac{i\omega t}{2}))
\end{split}
\end{equation}
for $\forall$ $\nu \in Z$.

Since  the replacement $-i\omega\rightarrow i\omega$ is equivalent to 
\begin{equation}
\begin{split}
\label{Eq42}
J_{\nu{'}} (\frac{2\sqrt{\frac{K_0}{m}}}{-i\omega}\exp(\frac{-i\omega t}{2}))\rightarrow
J_{-\nu}(\frac{2\sqrt{\frac{K_0}{m}}}{i\omega}\exp(\frac{i\omega t}{2})),
\end{split}
\end{equation}
 we obtain, that by $C^{[1]}_2 = 0$,  $C^{[2]}_2 = 0$  the expression (\ref{Eq33})
that is 
\begin{equation}
\begin{split}
\label{Eq43}
C^{[1]}_1  \exp(-\frac{\eta}{2m}t)J_\nu(\frac{2\sqrt{\frac{K_0}{m}}}{i\omega}\exp(\frac{i\omega t}{2})) 
\end{split}
\end{equation} 
will be the second linearly independent solution of the homogeneous damped Mathieu equation  $\forall C^{[1]}_{1} \neq 0 \in C$ and by $\forall\nu \in Z$ at $C^{[1]}_{1} = C^{[2]}_{1}$  for even $\nu \in Z$ and at $C^{[1]}_{1} = - C^{[2]}_{1}$  for uneven $\nu \in Z$, that is at $C^{[1]}_1 = C^{[2]}_1 sgn(-1)^{\nu}$. Its linear independence  from the first solution follows from linear independence of Bessel functions on the first and the second kinds.
At fixed $C^{[1]}_1$ and $C^{[1]}_2$ the relations  (\ref{Eq36}) and (\ref{Eq43}) represent themselves the fundamental system of the solution of homogeneous damped Mathieu equation. So, the general solution of homogeneous damped Mathieu equation is
\begin{equation}
\begin{split}
\label{Eq44}
&\exp(-\frac{\eta}{2m}t)[C^{[1]}_1  J_\nu(\frac{2\sqrt{\frac{K_0}{m}}}{i\omega}\exp(\frac{i\omega t}{2})) +\\
&C^{[1]}_2 Y_\nu(\frac{2\sqrt{\frac{K_0}{m}}}{i\omega}\exp(\frac{i\omega t}{2}))] 
\end{split}
\end{equation} 
The theorem is proved.

Thus, we have found the general solution of the  homogeneous damped Mathieu equation in the  essentially more simple form, allowing its practical using in many applications without numerical calculations, since Bessel functions are well known.

\end{document}